\documentclass[aps,prb,floatfix,twocolumn,superscriptaddress]{revtex4}
\usepackage{amsmath,amssymb}
\usepackage{graphicx}
\usepackage{epsfig}
\usepackage[usenames]{color}

\begin{document}


\title{Magnetoplasmon resonance technique to monitor\\two-dimensional Dirac superconductors in fluctuating regime}


\author{K.~Sonowal}
\affiliation{Center for Theoretical Physics of Complex Systems, Institute for Basic Science (IBS), Daejeon 34126, Korea}
\affiliation{Basic Science Program, Korea University of Science and Technology (UST), Daejeon 34113, Korea}

\author{V.~M.~Kovalev}
\affiliation{A.V.~Rzhanov Institute of Semiconductor Physics, Siberian Branch of Russian Academy of Sciences, Novosibirsk 630090, Russia}
\affiliation{Novosibirsk State Technical University, Novosibirsk 630073, Russia}

\author{I.~G.~Savenko}
\affiliation{Center for Theoretical Physics of Complex Systems, Institute for Basic Science (IBS), Daejeon 34126, Korea}
\affiliation{Basic Science Program, Korea University of Science and Technology (UST), Daejeon 34113, Korea}

\date{\today}

\begin{abstract}
We propose the magnetoplasmon resonance technique to investigate two-dimensional superconductors (taking MoS$_2$ as an example) in the fluctuating regime, where the temperature is slightly above the critical temperature of the superconducting transition. 
Thus, unpaired electrons and fluctuating Cooper pairs coexist in the system and interact with each other via long-range Coulomb forces, forming a Bose-Fermi mixture. 
We expose the sample to external time-dependent electromagnetic field with a frequency in sub-terahertz range and a permanent magnetic field, and show that the magnetoplasmon response of the system is strongly modified in the presence of superconducting fluctuations in the vicinity of the superconducting transition. 
In particular, the fluctuating Cooper pairs dramatically change the position and broadening of the magnetoplasmon resonance, which is reflected in the optical response of the system.
\end{abstract}

\maketitle


\section{Introduction}
The pioneering work of Bardeen, Cooper and Schrieffer (BCS) devoted to the theoretical explanation of the superconductivity phenomena~\cite{BCS} led to a sequential study of collective modes in superconductors~\cite{Anderson, Bogoliubov, VaksGalitskiiLarkin, CG1, CG2, CG3, SS, AV1, AV2}. 
It turned out that the plasmon modes are not at all affected by the transition of the material to the superconducting (SC) state since the nearly dispersionless plasmon frequency in any three-dimensional material is much larger than the SC gap. 
However, if the superconductor is in the peculiar \textit{fluctuating regime}~\cite{AL, Narlikar, VBLM, AV, Ketterson} at temperatures above than but close to the critical temperature of SC transition $T_c$, there start to emerge fluctuating Cooper pairs, but the value of the SC gap still remains zero. 
It seems essential to conclude, that in this situation there is no considerable influence of any SC properties on the plasma oscillations of normal electrons in the sample. 
Indeed, the correction to the Drude conductivity caused by the SC fluctuations, known as paraconductivity, is as a rule small~\cite{LarkinVarlamov} albeit observable~\cite{RefReviewGlover}.
Yet, let us not jump to conclusions. 
Owing to the presence of SC fluctuations, a lot of intriguing physics occurs in superconductors above $T_c$\cite{Kumar,Shai,Bossoni_2014}, and this is of great interest specially in low-dimensional\cite{RefOurAL, RefCPE} and high $T_c$ superconductors\cite{Rullier,Perfetti}. 

Experimentally, the study of plasmons is usually based on the measurement of the electromagnetic (EM) power absorption by the sample, which is proportional to the electric current in the system. 
It is true that the transport of normal electrons gives the dominant contribution to the EM power absorption, and the impact of the paraconductivity terms is small.
However, the fluctuations can essentially influence both the \textit{position} and \textit{damping} of the plasmon resonance of normal electrons~\cite{RefOurAL}.
It stems from strong interaction between normal electrons and the fluctuating Cooper pairs in the sample via the long-range Coulomb forces, which are activated while exciting the plasmons with an EM field. 

The position of the plasmon resonance is also sensitive to the presence of an external magnetic field. 
It gives an opportunity to control the magnetoplasmon (MP) resonance (MPR) in the experiment. 
In this paper, we develop a theory of MPR in two-dimensional superconductors in the fluctuating regime (at temperatures $0<T-T_c\ll T_c$). 
The system we study represents a single layer of a two-dimensional (2D) material at the temperature slightly above the $T_c$, exposed to an electromagnetic field and an external magnetic field. 
Thus, the system represents a mixture of a Fermi gas of normal electrons and a Bose gas of fluctuating Cooper pairs interacting via the Coulomb forces.

Other examples of hybrid Bose-Fermi systems comprise electron gas - superconductor~\cite{BKS2, RefOurKus, RefOurShed}, exciton gas - 2D electron gas (2DEG)~\cite{Villegas}, and exciton-polaritons in a microcavity - 2DEG systems~\cite{Shelykh}. 
From that perspective, a superconductor in the fluctuating regime can be looked at as a hybrid Bose-Fermi system, where two gases described by different quantum statistics coexist in one (same) 2D layer, forming a Bose-Fermi mixture. 

The MPR has already been studied in some hybrid systems, in particular, a system consisting of a degenerate electron gas interacting  with a dipolar exciton gas in the Bose-Einstein condensate (BEC) state has been considered~\cite{BKS}.
There, the theoretical description of excitonic BEC in the framework of the Bogoliubov model demonstrates the existence of sound-like collective modes of the BEC. 
The phase velocity of these modes dramatically depends on the condensate  density and can strongly interact with plasmon modes of electronic subsystem. 
This interaction modifies the MPR of the whole system resulting in the replacement of the standard Lorenz shape of MPR by an asymmetric profile which reminds of Fano resonance~\cite{RefRMPFano}.
Qualitatively, this behavior can be explained by a simple harmonic oscillator model. 
Indeed, the two oscillatory modes (plasmon oscillation of the electron density and sound-like modes of excitonic BEC) can be looked at as harmonic oscillators coupled with each other by a spring (the Coulomb force) and exposed to an external driving (EM field). 
One thus have all the necessary ingredients to observe the (MP) Fano resonance phenomenon~\cite{RefRMPFano}.

In the case of fluctuating superconductors (where the bosonic subsystem is fluctuating Cooper pairs but not the BEC particles as in the example before), the situation is different. 
The Bose gas of fluctuating Cooper pairs does not possess its own collective modes. 
Thus, when an external EM field excites the plasmon oscillations of normal electron gas subsystem, the Bose subsystem (SC fluctuations) does not oscillate since the plasmon phase velocity exceeds both the Fermi velocity and the phase velocity of fluctuating Cooper pairs. Therefore, the gas of SC fluctuations plays the role of viscous media for plasmon oscillations of normal electrons.
It results in the appearance of frictional force acting on the plasmon harmonic oscillator. 
As a result, in the vicinity of the critical temperature $T_c$, the shape of the plasmon resonance experiences strong modification. 

%
%
\begin{figure}
\includegraphics[width=0.5\textwidth]{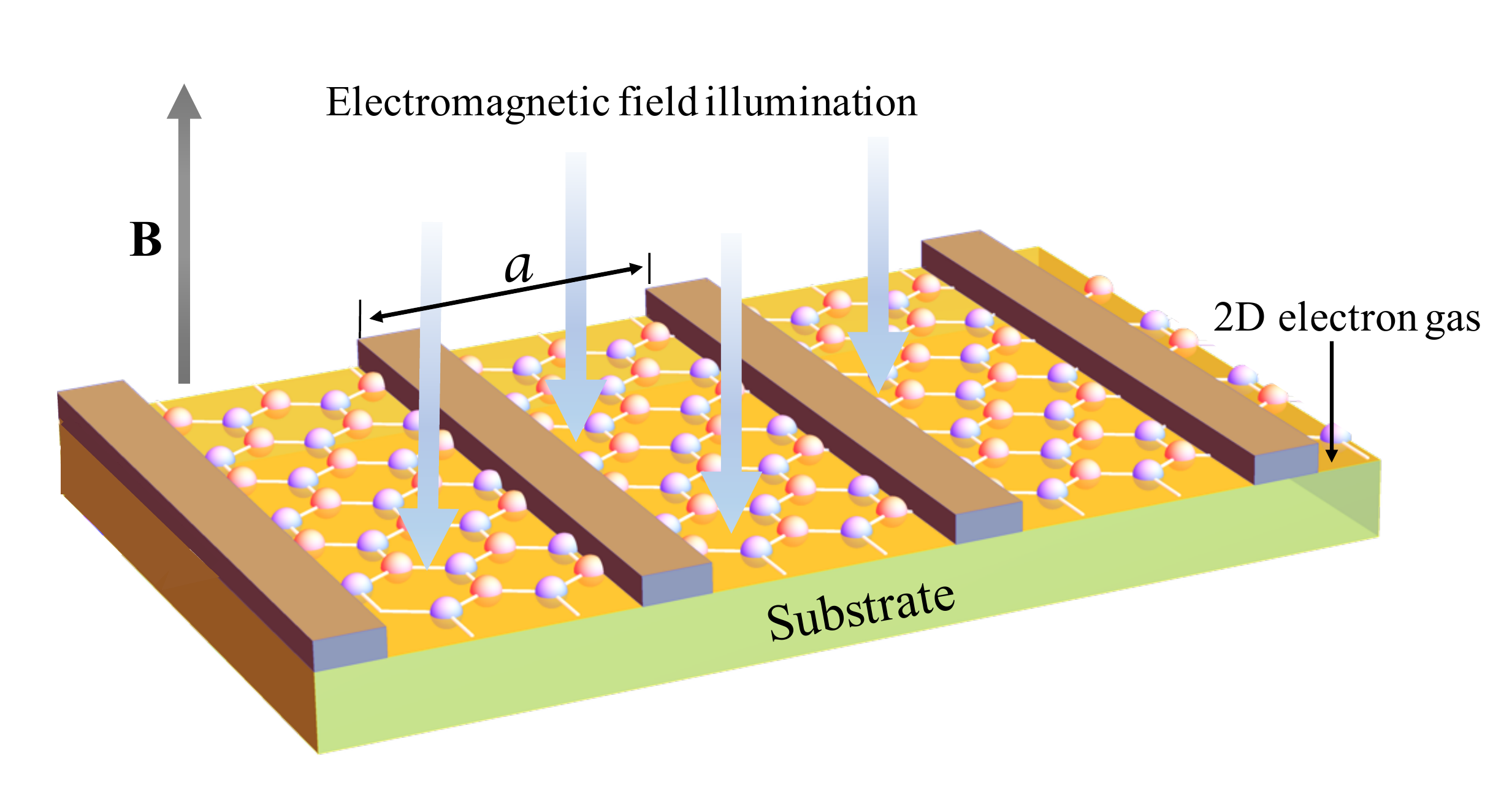}
\caption{System schematic. A two-dimensional electron gas covered by a metallic grating. The system is exposed to an EM  field of light $\mathbf{E}$ polarized across the grating and a uniform magnetic field $\mathbf{B}$.}
\label{Fig1}
\end{figure}
%
%
%


\section{theory}

We consider a system that resembles a standard setup for the experimental investigation of the MPR phenomenon (Fig.~\ref{Fig1}). 
It consists of a 2D conducting layer covered with a metallic grating. 
As an example, we use a MoS$_2$ monolayer, where the SC transition has been recently observed experimentally at high electron densities \cite{Wakatsukie,Saito}. An external EM field creates a spatially-modulated effective field acting on the 2D electrons. 
The grating period $a$ determines the wave vector $k=2\pi/a$ of the effective EM field. 
Furthermore, the system is additionally exposed to a permanent magnetic field.
Magnetoplasmons have been widely studied in two-dimensional electron systems using different methods \cite{Dahl,Deviatov,Lee,Jingbo}, paving the way for variety of applications\cite{plasmonics,Padmanabhan}. 
To study the magnetoplasmons (MPs) in the presence of SC fluctuations, we will use the approach based on the analysis of the dielectric function. 

For the 2D layer placed on top of the substrate at $z>0$, the Poisson equation for the scalar potential $\phi(z)$ due to the internal induced electric field reads
\begin{equation}
\label{Poisson}
    \Bigg(\frac{\partial}{\partial z} \kappa(z) \frac{\partial}{\partial z} - k^2\Bigg) \phi(z) = -4 \pi (\varrho_e + \varrho_{AL})\delta(z),
\end{equation}
where $\varrho_e$ and $\varrho_{AL}$ are the charge densities of normal electrons and fluctuating Cooper pairs in reciprocal space, respectively. 
We note that while solving the Poisson equation, we neglect the retardation effects.
Using the solution of Eq.~\eqref{Poisson} and the continuity equation, we arrive at the dielectric function given by
\begin{gather}
\label{DF}
\varepsilon(\textbf{k},\omega)=1+i\frac{4\pi k}{(\kappa+1)\omega}\Bigl[\sigma_{xx}(\textbf{k},\omega)+\varrho_{xx}(\textbf{k},\omega)\Bigr],
\end{gather}
where $\kappa$ is the dielectric constant. The equation which defines the dispersion of collective modes in the system reads~\cite{vitlinachaplik, sarma, chapliknew}
\begin{gather}
\label{EQ1}
\varepsilon(\textbf{k},\omega)=0,
\end{gather}
where $\sigma_{xx}(\textbf{k},\omega)$ is the longitudinal part of the Drude magnetoconductivity tensor and $\varrho_{xx}(\textbf{k},\omega)$ is the longitudinal part of paraconductivity in a uniform magnetic field $B$. 
The phase velocity of the plasmon mode is much larger than the Fermi velocity of electrons and it also exceeds the center-of-mass velocity of Cooper pairs. 
This allows us to neglect the spatial dispersions of both the conductivities, thus assuming $\textbf{k}=0$ in their  expressions.

The Drude conductivity of normal electrons has its conventional form,
\begin{gather}
\label{EQ2}
\sigma_{xx}(\omega)=\sigma_0\frac{i(i+\omega\tau)}{(i+\omega\tau)^2-(\omega_c\tau)^2},
\end{gather}
where $\sigma_0=e^2n\tau/m$ is the static Drude conductivity of normal electrons with $e$ the electric charge, $n$ the electron density, $\tau$ the scattering time, and $m$ the effective electron mass; $\omega_c=eB/m$ is the cyclotron frequency. 
The experimental observation of the plasmon resonance holds under the condition $\omega\tau\gg1$, thus preventing the impurity-induced plasmon damping. 
A typical range of plasmon frequencies is $\omega\sim\,10^{10} \div 10^{11}$~s$^{-1}$, and for $T_c\sim 10$~K we estimate $\omega/T_c\sim10^{-2}\div10^{-1}$. 
Thus, we find (for typical parameters) $T_c\gg\omega\gg1/\tau$. 
Under this condition, the superconductor is in the \textit{clean limit}, where the nonlocality effects play an essential role \cite{Aronov}. 
The general expression for paracoductivity at finite frequency, magnetic field, and non-locality corrections reads~\cite{Aronov}
\begin{gather}\label{para}
\varrho_{xx}(\omega,\omega_c)=\varrho^0_{xx}(\omega,\omega_c)\left[\frac{1-i\omega\tau+2\omega_c^2\tau^2}{(1-i\omega\tau)^2+\omega_c^2\tau^2}\right]^2,
\end{gather}
where $\varrho^0_{xx}(\omega,\omega_c)$ is the paraconductivity in `local approximation', and the remaining part (the terms in square brackets) accounts for the non-local corrections. Taking into account that $T_c\gg\omega$, the expression for $\varrho^0_{xx}(\omega,\omega_c)$ can be written in the static limit $\omega=0$, 
\begin{gather}
\label{EQ3}
\varrho^0_{xx}(\omega_c)=\varrho_0\mathcal{F}\left(\frac{\mu}{\omega_c}\right),\\\nonumber
\mathcal{F}(x)=8x^2\left[\psi\left(\frac{1}{2}+x\right)-\psi(x)-\frac{1}{2x}\right],
\end{gather}
where $\varrho_0=e^2/(16\epsilon)$ is the static Aslamazov-Larkin (AL) conductivity of fluctuating Cooper pairs. 
In what follows, we will (naturally) assume the smallness of AL correction as compared with the static Drude conductivity, $\varrho_0/\sigma_0\ll1$. In Eq.~\eqref{EQ3}, $\mu=\alpha T_c\varepsilon$, where $\varepsilon=(T-T_c)/T_c$ is the reduced temperature, and the Ginzbug-Landau parameter $\alpha$ is fixed by the relation $4m\alpha T_c\xi^2=1$. The coherence length $\xi$ in 2D samples reads
\begin{gather}\label{EQ11}
\xi^2=\frac{v_F^2\tau^2}{2}\Bigl[
\psi\left(\frac{1}{2}\right)
-
\psi\left(\frac{1}{2}+\frac{\hbar}{4\pi T\tau}\right)
+\frac{\hbar\psi'\left(\frac{1}{2}\right)}{4\pi T\tau}\Bigr],
\end{gather}
where $\psi(x)$ is the digamma function and $v_F$ is the Fermi velocity. 
\begin{figure}[!t]
\includegraphics[width=0.45\textwidth]{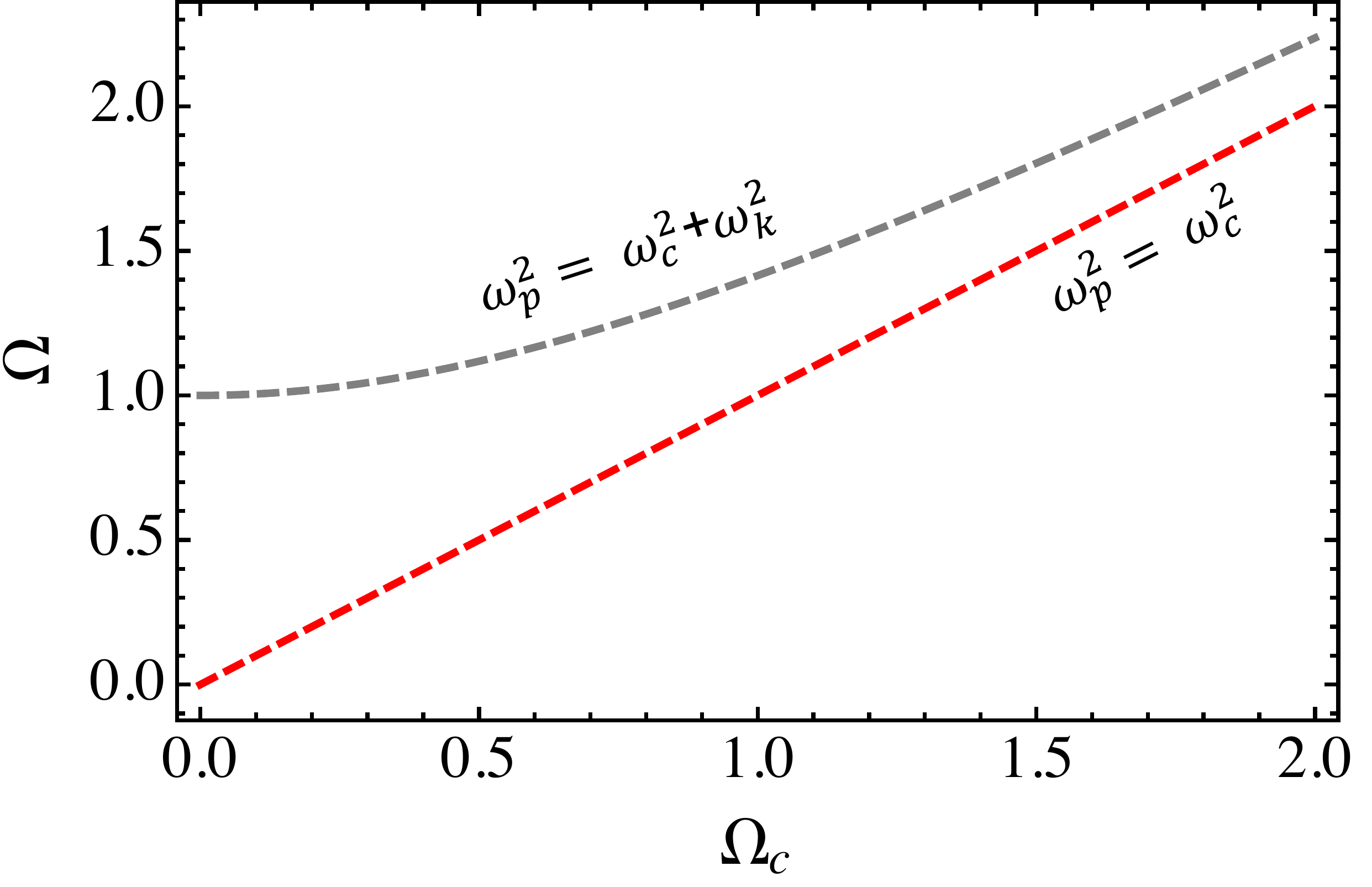}
\caption{MPR dispersions without the account of the SC fluctuations.
Gray and red dashed lines show the asymptotic behavior  $\omega_p=\sqrt{\omega_k^2+\omega_c^2}$ and $\omega_p=\omega_c$, respectively.  }
\label{Fig2}
\end{figure}

In the absence of SC fluctuations, using~\eqref{EQ2}, we come up with the dispersion equation 
%
\begin{gather}
\label{EQ4}
\frac{\omega_k^2}{\omega}\frac{\omega+i/\tau}{(\omega+i/\tau)^2-\omega_c^2}=1,
\end{gather}
where we introduce 2D bare plasmon frequency $\omega_k=\sqrt{4\pi e^2nk/(\kappa+1)m}$. In the limit of weak electron-impurity collisions, $\omega\tau$, $\omega_c\tau\gg1$, the MP dispersion reads
\begin{gather}
\label{EQ5}
\omega_p=\sqrt{\omega_k^2+\omega_c^2}.
\end{gather}
This holds if $\omega_p\tau\gg1$, which is a typical condition for experimental observation of the MPR. 
Expression~\eqref{EQ5} shows that at vanishing magnetic fields, the MP dispersion coincides with bare 2D plasmon frequncy $\omega_k$, whereas with increasing magnetic field, $\omega_p$ approaches the cyclotron frequency $\omega_c$, as it is shown in Fig.~\ref{Fig2}.
To build the plots, we introduce the dimensionless quantities $\Omega=\omega/\omega_k$ and $\Omega_c=\omega_c/\omega_k$.
In order to describe the actual experiment, we use the parameters typical for $\rm {MoS_2}$~\cite{Kukushkin}: $n\sim 10^{11}$~cm$^{-2}$,  $\omega_p\sim10^{10}$~s$^{-1}$.
If larger frequencies are required, one can use the electron density up to $n\sim 10^{14}$~cm$^{-2}$, which has recently been reported in MoS$_2$~\cite{Wakatsukie}. 

\begin{figure*}
\includegraphics[width=0.9\textwidth]{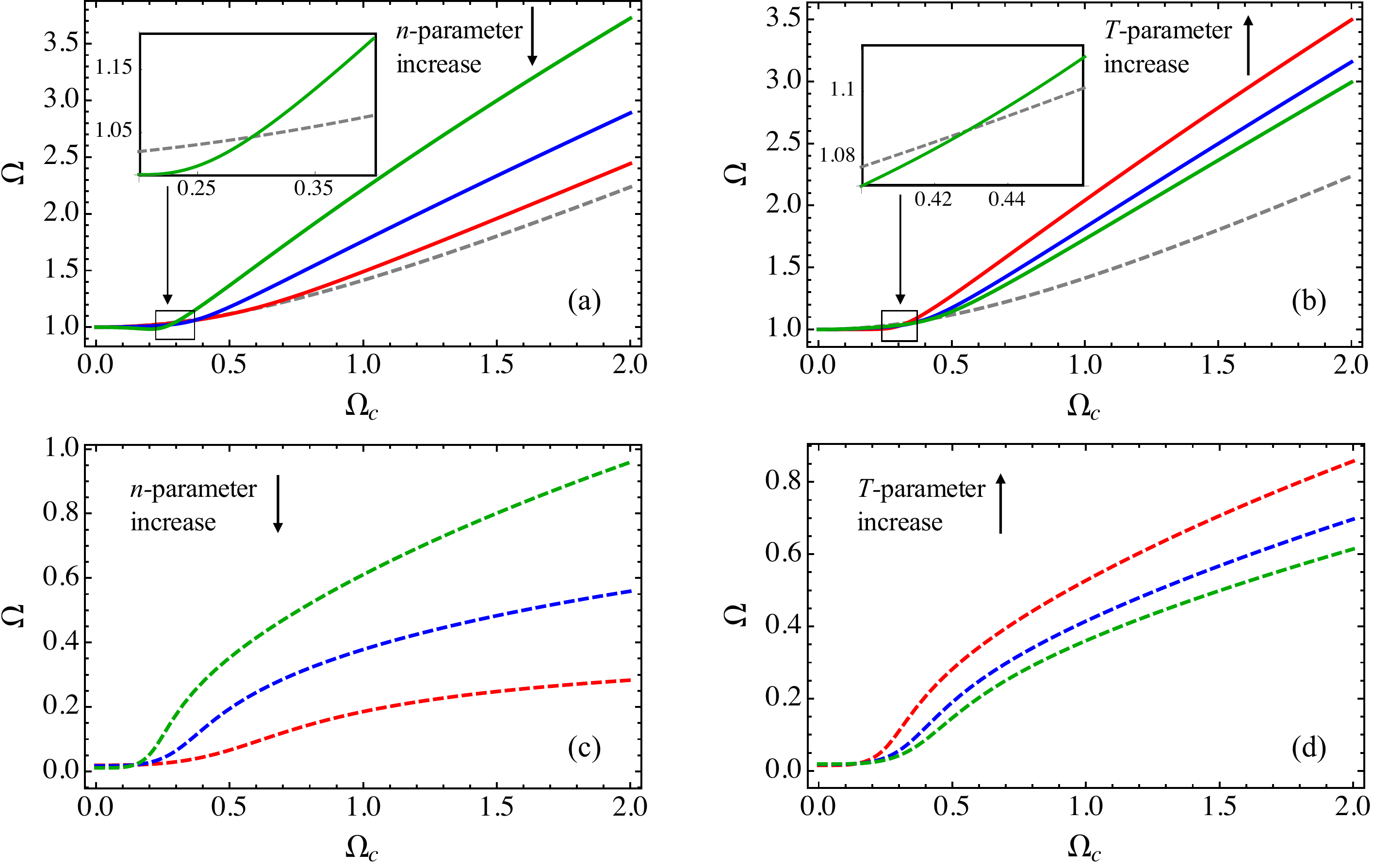}
\caption{Real [(a) and (b)] and imaginary [(c) and (d)] parts of the MPR dispersions. 
Panels (a) and (c) show the dependence of the MPR dispersion on the electron density: $n=1.5 \times 10^{11} {\rm cm^2}$ (green), $5 \times 10^{11} {\rm cm^2}$ (blue), and $1.5 \times 10^{12} {\rm cm^2}$ (red) at $T = 10.1$~K. 
Panels (b) and (d) show the dependence of the MPR on temperature: $T = 10.08 $~K  (red), $10.05$~K (blue), and $10.02 $~K (green) at $n=1.5 \times 10^{11} {\rm cm^2}$.}
\label{Fig3}
\end{figure*}
%
%
%

%
%
%




To find the MP dispersion in the presence of SC fluctuations, we should consider the dispersion equation~\eqref{EQ1} together with the conductivities, given by Eqs.~\eqref{EQ2} and~\eqref{para} (the renormalization of the MP dispersion in general case will be analyzed numerically below). 
Let us estimate here analytically the MP damping due to SC fluctuations in the limit of weak electron-impurity scattering, $\omega_p\tau\gg1$. 
After some algebra, the dispersion equation Eq.~\eqref{EQ2} takes the form
\begin{gather}
\label{EQ6}
\omega^2-\omega_p^2=\frac{1-2i\omega\tau}{\tau^2}+i\frac{\omega_k^2}{\omega\tau}
\left[1+\eta_c\frac{(1-i\omega\tau+2\omega_c^2\tau^2)^2}{(1-i\omega\tau)^2+\omega_c^2\tau^2}\right],
\end{gather}
where
\begin{gather}
\label{EQ7}
\eta_c=\frac{\varrho_0}{\sigma_0}\mathcal{F}\left(\frac{\mu}{\omega_c}\right)
\end{gather}
determines the strength of interaction between the normal electrons and SC fluctuations. 

Eq.~\eqref{EQ6} is still exact. 
In the limit $\omega\tau\gg1$, it can be solved by successive approximations. 
In the lowest order, we substitute $\omega=\omega_p$ in the right-hand side of this expression, and from Eq.~\eqref{EQ6} we find $\omega=\omega_p-i\Gamma$, where the MP damping reads
\begin{gather}
\label{EQ10}
\Gamma=\frac{1}{\tau}-\frac{\omega_k^2}{2\omega_p^2\tau}
\left[1-\eta_c\frac{4\omega_c^4\tau^2-\omega_p^2}{\omega_k^2}\right].
\end{gather}
This analytical expression accounts for both the electron-impurity scattering and the influence of SC fluctuations to the MP damping (in the limit $\omega_p\tau\gg1$).  
In the absence of magnetic field ($\omega_c=0$) we find
\begin{gather}
\label{EQ10.2}
\Gamma=\frac{1-\varrho_0/\sigma_0}{2\tau}.
\end{gather}
Comparing it with the dispersion of the 2D plasmon ($\omega=\omega_k-i/2\tau$), we conclude that the presence of SC fluctuations results in the \textit{narrowing} of the resonance since $\varrho_0/\sigma_0>0$, and thus $1-\varrho_0/\sigma_0<1$. In the case of large magnetic fields, when $\omega_c\gg\omega_k$, we have
\begin{gather}
\label{EQ10.3}
\Gamma=\frac{1+2(\omega_c\tau)^2\eta_c}{\tau}.
\end{gather}

It should be noted, that the function $\mathcal{F}$ in Eq.~\eqref{EQ3} is limited, $0<\mathcal{F}\left(\mu/\omega_c\right)<1$. 
Thus, the value of $\eta_c$ [in Eq.~\eqref{EQ10.3}] is determined mainly by the prefactor $\varrho_0/\sigma_0$. 
Despite the fact that the AL theory is applicable at $\varrho_0/\sigma_0\ll1$, this smallness can be compensated by the `plasmonic factor' $\omega_c\tau\gg1$, which linearly grows with $B$. 
Thus, strong magnetic fields may amplify the SC fluctuations contribution into the MP damping. 
Contrast to the zero-field limit, at large magnetic fields, the SC fluctuations result in the \textit{broadening} of the width of MPR.  



\section{ results}

To study the MP dispersion numerically (at arbitrary parameters), it is convenient to rewrite Eq.~\eqref{EQ6} in terms of dimensionless parameters $\Omega$ and $\Omega_c$ (introduced before). 
Figure~\ref{Fig3} shows the MP dispersion at different values of normal electron densities and reduced temperature, calculated numerically according to Eq.~\eqref{EQ1}. 
Both figures demonstrate an essential modification of both the position (which corresponds to real part of the eigen energy) and width (damping) of MPR. 
For comparison, we also show the gray dashed curves (in both the figures), which stand for the MP dispersion in the absence of SC fluctuations (compare also with Fig.~\ref{Fig2}). 

The MPR exists for all the values of magnetic field. 
In the case of zero magnetic field, the MPR dispersion resembles the one without the SC fluctuations. 
Obviously, the renormalization of the MPR due to the presence of SC fluctuations becomes more pronounced as temperature approaches the critical temperature $T_c$, in the vicinity of which the density of SC fluctuations increases. 
In contrast, an increase of normal electron density suppresses the SC fluctuations contribution to both the MPR damping and dispersion.



Another interesting feature is that the MPR dispersion renormalization is non-monotonous, as it is seen in Fig.~\ref{Fig3}(a) and Fig.~\ref{Fig3}(b). 
At weak magnetic fields, it demonstrates a redshift with respect to the bare MP dispersion, whereas at large fields it acquires a blueshift. 
Figure~\ref{Fig3}(c) shows the variation of MPR damping with electron density. 
We note, that the magnitude of damping decreases and the peaks get broader with the increase of the electron density. 
From Fig.~\ref{Fig3}(d) we can conclude that as we increase the temperature, the MPR is gets damped. 
An increase in temperature implies that the proximity to $T_c$ is lowered, and hence the influence of SC fluctuations also decreases.

\section{Conclusions}
In conclusion, we have studied the dielectric function of a two-dimensional material in the regime of superconducting fluctuations (in the vicinity of the transition temperature to the superconducting state from above), exposed to an electromagnetic field of sub-terahertz frequency and a constant magnetic field.
We analyzed the dispersion of collective modes in the system, taking into account the interaction between the normal electrons and fluctuating Cooper pairs influenced by the magnetic field. 

This paper sheds light on the magnetoplasmon spectroscopy (which represents a well-established experimental technique) of superconductors in fluctuating regime.
This technique can be especially effective in samples of reduced dimensionality, since the paraconductivity correction becomes more significant in lower-dimensional structures~\cite{LarkinVarlamov}; it can also be employed to monitor multi-layer structures based on transition metal dichalcogenides or cuprates.
For that, one has to modify Eq.~\eqref{DF} for the dielectric function and account for the paraconductivity term in the dielectric function for layered materials~\cite{Mahan, Artemenko, Slipchenko}.


\acknowledgements
We acknowledge the support by the Institute for Basic Science in Korea (Project No.~IBS-R024-D1), the Russian Foundation for Basic Research (Project No.~18-29-20033), and the Ministry of Science and Higher Education of the Russian Federation (Project FSUN-2020-0004).


\bibliography{mpr.bib}

\end{document}